\documentclass[5p,twocolumn,numbers]{elsarticle}
\usepackage{newtxtext,newtxmath}
\usepackage[utf8]{inputenc}
\usepackage{url}
\usepackage{hyperref}
\usepackage{graphicx}
\usepackage{booktabs}
\usepackage{amsmath}
\usepackage{comment}
\usepackage{makecell}
\usepackage{tikz}
\usetikzlibrary{arrows.meta,positioning,calc,fit,backgrounds}
\usepackage{orcidlink}
\begin{document}

\begin{frontmatter}

\title{Trustless Accountable Data Sharing for Supply Chains: A Reference Architecture and MoreMedDiet Proof-of-Concept}

\author[inst1]{Michal Kit \orcidlink{0009-0000-9105-0422}}
\author[inst1]{Montassar Naghmouchi \orcidlink{0000-0003-3467-7514}}
\author[inst1]{Maryline Laurent \orcidlink{0000-0002-7256-3721}\corref{cor1}}
\ead{maryline.laurent@telecom-sudparis.eu}
\author[inst1]{Badis Hammi \orcidlink{0000-0002-4470-6406}}
\author[inst2]{Hella Kaffel Ben Ayed \orcidlink{0000-0002-4433-7719}}
\author[inst1]{Mohamed Amine Hamdi}
\author[inst2]{Jihene Khoualdi\orcidlink{0009-0004-5614-5636}}
\author[inst2]{Ilhem Abdelhedi Abdelmoula\orcidlink{0000-0001-5866-7252}}
\author[inst3]{Sami Sboui}

\cortext[cor1]{Corresponding author}

\affiliation[inst1]{organization={SAMOVAR, Télécom SudParis, Institut Polytechnique de Paris},
    city={Palaiseau},
    country={France}}

\affiliation[inst2]{organization={LIPAH, Faculty of Sciences of Tunis, University of Tunis El Manar},
    city={Tunis},
    country={Tunisia}}

\affiliation[inst3]{organization={JASSP SAS},
    city={Paris},
    country={France}}


\begin{abstract}
Modern supply chains are increasingly dependent on the exchange of sensitive business data between heterogeneous stakeholders who do not trust each other. 
Conventional integration approaches typically rely on centralized intermediaries or require extensive data disclosure, which hinders collaboration, compliance, and innovation. We propose a generic reference architecture for privacy preserving, trustless data sharing in supply chains. The architecture combines permissioned distributed ledgers, IoT-based evidence ingestion, and modular privacy mechanisms to enable controlled transparency: stakeholders can extract verifiable insights and support audits while minimizing data exposure. 

We instantiate the architecture in the context of the MoreMedDiet project, which promotes sustainable farm-to-fork food systems and adoption of the Mediterranean diet. By mapping the MoreMedDiet use case into our architecture, we demonstrate how channel topologies, IoT-to-ledger data flows, and smart contract-based quality assurance (QA) can support secure data exchange, transparency, interoperability, and regulatory alignment. Our proof-of-concept illustrates how Self-Sovereign Identity (SSI) and W3C Verifiable Credentials (VCs) can provide portable and auditable access governance over Hyperledger Fabric channels. 

The resulting design offers a reusable blueprint for multi-actor supply chains, including agri-food and pharmaceutical domains. Moreover, the performance achieved -- presenting a verified credential to the next stakeholder in the supply chain in 130 ms -- is promising and demonstrates the potential applicability of the approach in real-world supply chain environments.
\end{abstract}


\end{frontmatter}

\section{Introduction}

Regulatory and societal pressures are pushing supply chains towards greater transparency, sustainability, and ethical accountability. In the European Union, the Digital Product Passport (DPP) \cite{DPP} requires products to carry digital identifiers (e.g.\ QR codes, RFID tags or barcode) that link stakeholders and consumers to verifiable information about product's origin, production methods, and lifecycle impact. While such measures promise unprecedented visibility, they also present fundamental challenges regarding interoperability, privacy, and trust in heterogeneous and competitive ecosystems \cite{Ozturk}.

Food supply chains epitomize these challenges. Producers, transporters, processors, distributors, retailers, and regulatory authorities must exchange data that is subject to strict traceability and quality assurance protocols. However, these exchanges occur between parties with divergent incentives and limited mutual trust, which makes achieving transparency without overexposing commercially sensitive or personal data difficult.

Emerging technologies---notably permissioned blockchains, the Internet of Things (IoT), and smart contracts---offer new ways to address these issues. IoT sensing enables the continuous monitoring of product conditions, while distributed ledgers provide tamper-evident and auditable records, as well as consensus across organizations. Smart contracts support automated quality checks, compliance verification, and the enforcement of business rules, eliminating the need for centralized intermediaries.

In this work, we present a reference architecture for trustless, privacy preserving sharing of data in multi-actor supply chains. This architecture incorporates a permissioned ledger, IoT-based evidence ingestion, and on-/off-chain data placement, providing flexible levels of transparency alongside strong integrity and accountability guarantees. To demonstrate its applicability, we instantiate the architecture in the MoreMedDiet (More on the adoption of a healthy Mediterranean diet) project \cite{MMD}, which aims to promote the Mediterranean diet through a sustainable farm-to-fork supply chain. The MoreMedDiet scenario illustrates the convergence of traceability, live monitoring, quality assurance, and controlled data disclosure requirements.

\textbf{Contributions.} This work makes four main contributions. First, we propose a generic reference architecture for controlled transparency in multi-actor supply chains. This architecture is based on a permissioned ledger, IoT-based evidence ingestion, and on-/off-chain data partitioning. Second, we design and implement an SSI/VC-based access governance module in which an issuer service provides verifiable role, mandate, and certification credentials, while applications or gateways verify verifiable presentations at the request boundary. This enables portable and auditable authorization across organizations
. Please note that securing IoT devices and gateways is outside the scope of this contribution, as our focus is on the supply chain model and traceability.
Third, we present a proof-of-concept for the MoreMedDiet project. This demonstrates the channel topology, the IoT-to-ledger data flow, smart contract-based quality assurance (QA) checks, and auditor workflows in a realistic agri-food scenario. The PoC demonstrates the feasibility of our proposal and the potential applicability of our model to real-world supply chain environments, as shown by the performance results achieved.
Fourth, in addition to the functional and regulatory requirements, we pay particular attention to security and privacy requirements. We compile these requirements carefully and demonstrate how our architecture meets them. 

The remainder of this paper is organized as follows. Section~\ref{sec:background} summarizes relevant background and related work. Section~\ref{sec:requirements} introduces the organizational system model, along with the functional, security, privacy, and threat assumptions arising from the participating stakeholders. Section~\ref{sec:architecture} then presents the technical reference architecture, including its technical system model and complementary technical requirements and threat assumptions. Section~\ref{sec:security} analyzes the security and privacy properties of the generic architecture. Section~\ref{sec:implementation} describes the MoreMedDiet proof-of-concept implementation, providing performance measurements and analysis to demonstrate the model's potential applicability in real-world supply chain environments. Section~\ref{sec:conclusion} concludes and outlines future work.

\section{Background and Related Work}
\label{sec:background}

This section provides a brief overview of the enabling technologies and prior work that our architecture is based on, including permissioned blockchains, Self-Sovereign Identity (SSI) and Verifiable Credentials (VCs), IoT-enabled supply chains, and related work on privacy preserving data sharing across organizational boundaries.

\subsection{Blockchain and Hyperledger Fabric in Supply Chains}

Blockchain technology provides a shared, tamper-evident ledger for recording transactions and state changes across organizational boundaries. In supply chains, blockchain can reduce information asymmetries by establishing a common, auditable view of product provenance, handovers, and quality-related events. Integrating smart contracts with IoT data further supports automated QA checks, event-triggered actions, and conditional payments.

Public blockchains, however, are often ill-suited to enterprise supply chains due to scalability, confidentiality, and governance constraints. In such settings, private, permissioned ledgers offer a better fit. Hyperledger Fabric (HF) \cite{HF} is a prominent permissioned blockchain framework designed for networks of known organizations requiring confidentiality, accountability, and deterministic execution. Its modular architecture separates transaction execution, ordering, and validation, enabling flexible consensus and policy configuration, and supporting high throughput in enterprise consortia.

A key feature of Fabric for supply-chain contexts is its channel abstraction \cite{priv-channel}. Channels define logically isolated ledgers within a single network; each channel maintains its own chaincode and transaction history visible only to authorized members. Fabric also supports private data collections, which replicate only hashes of confidential data on the ledger while keeping plaintext confined to permitted peers. Identity and access management rely on a Membership Service Provider (MSP) that uses X.509 certificates to authenticate actors and enforce channel policies. These features make Fabric suitable for multi-actor supply chains with partial trust, as they facilitate selective transparency, fine-grained access control, and auditable collaboration.

\subsection{Self-Sovereign Identity and Verifiable Credentials}
\label{subsec:ssi_vc}

Self-Sovereign Identity (SSI) is an approach to digital identity in which entities can hold and present cryptographically verifiable claims about themselves without relying exclusively on a single centralized identity provider. Instead of requiring each organization to manage access through isolated identity silos, SSI enables portable identity assertions that can be issued by one party, held by another, and verified by many relying parties across organizational boundaries.

A central concept in SSI is the \emph{Decentralized Identifier} (DID), standardized by the W3C \cite{DID}. A DID is a globally unique identifier associated with cryptographic material and metadata that allow an entity to prove control over that identifier. In practice, DIDs can represent organizations, individuals, devices, or software services. Unlike conventional identifiers tied directly to a specific provider, DIDs are intended to support decentralized resolution and control.

A second key concept is the \emph{Verifiable Credential} (VC) \cite{VC}, also standardized by the W3C. A VC is a cryptographically signed statement issued by an authority about a subject. In the context of supply chains, a VC may attest properties such as organizational role, certification status, audit mandate, device registration, or authorization to access specific data or services. The party that issues the credential is the \emph{issuer}, the party that holds it is the \emph{holder}, and the party that checks it is the \emph{verifier}.

To use a VC in an interaction, the holder typically generates a \emph{verifiable presentation} (VP), which contains one or more credentials or selectively disclosed attributes derived from them, together with proof that the presenter controls the corresponding identifiers. This allows a verifier to check whether the required conditions are satisfied without necessarily learning all information contained in the original credentials. Such selective information disclosure is particularly attractive in regulated supply chains, where participants often need to prove eligibility, role, or mandate without revealing unrelated internal information.

In our architecture, SSI complements rather than replaces the baseline identity mechanisms of Hyperledger Fabric. Fabric MSPs continue to provide network-level membership and authentication for the permissioned ledger, while SSI and VCs provide a portable and application-layer mechanism for expressing roles, certifications, mandates, and access rights across organizations. This separation allows the system to combine ledger-native trust and governance with more flexible and privacy preserving authorization at the boundaries of applications, gateways, and off-chain services.

\subsection{IoT Devices in Supply Chains}

The Internet of Things has transformed data acquisition and monitoring in modern supply chains \cite{IoT-sup}. Sensors, actuators, and gateways collect and transmit data about environmental conditions, location, and process parameters. In the agri-food sector \cite{agri-chain}, IoT sensors support continuous monitoring from farm to fork, enabling automated alerts, predictive maintenance, and data-driven optimization of resource usage.

IoT-enabled supply chains offer real-time monitoring and automated logging of product and process conditions, enhanced traceability and evidence for QA and certification, and opportunities for sustainability improvements through fine-grained tracking of resource and waste indicators. At the same time, they must cope with device heterogeneity, interoperability challenges, security vulnerabilities of resource-constrained devices, and the difficulty of ensuring trustworthy measurements in competitive environments. Large-scale deployments must manage high data volumes without overwhelming storage or network capacity.

In our architecture, IoT devices serve as data providers for the ledger. Gateways aggregate measurements, perform basic validation, and submit evidence to appropriate channels, with design choices balancing data granularity, scalability, and privacy. This IoT-to-ledger pattern becomes a central mechanism for linking physical-world observations to digital evidence.

\subsection{Related Work}

Prior work has extensively studied permissioned blockchains for supply-chain traceability and IoT integration. Comparative evaluations of permissioned frameworks highlight Hyperledger Fabric's strengths in performance, scalability, and channel-based data isolation for enterprise collaboration \cite{polge}. In \cite{Khan2022} an empirical baseline for HF in supply chains is established. HF demonstrated consistent throughput during high-volumes of simultaneous transactions - between 500 and 1000 transactions per second (TPS) - and good scalability up to 20 nodes where latency actually decreases. Apart from performance, HF comes with native channel isolation, but it lacks the fine-grained and portable access control, highly desired in dynamic regulatory compliance.
Other work demonstrates the ingestion and management of IoT data with Fabric, including chaincode and data-lifecycle design for sensor telemetry. Decentralized access-control designs for IoT using public blockchains and off-chain data storage have also been explored~\cite{shafagh}, foreshadowing patterns that separate control and data planes for scalability and privacy. Sector-specific studies, for example, in construction supply chains, show how distributed ledgers can enhance provenance and tamper-evidence in multi-tier supplier networks. 

However, most existing solutions neglect the need for a decentralized identity layer that complements the blockchain traceability with privacy for supply chain data exchange between actors. Indeed, the balance between transparent traceability and operational confidentiality is a fundamental requirement for multi-actor supply chains. Systematic reviews have shown that while blockchains provide better data integrity and permanent immutable audit trails, integration of privacy preserving mechanisms remains a primary challenge \cite{Ellahi2024}.

Recent architectures have been proposed to bridge this gap utilizing secondary privacy mechanisms onto blockchain ledgers. For example, \cite{Keo2025} utilizes differential privacy (DP) with HF and IPFS (InterPlanetary File System) for off-chain storage, highlighting the necessary trade-off between transaction obfuscation and its efficiency. Other frameworks like PrivChain \cite{Malik2022} utilize client side Zero-Knowledge Range Proofs (ZKRPs) to selectively disclose information from the supply chain without revealing exact provenance. The proposed framework also allows for off-line proof computation and commitments, minimizing the overhead of computation over blockchain. While Zero-Knowledge Proofs (ZKPs) preserve business confidentiality, this cryptographic privacy incurs a computational overhead causing the ZKRP transactions latency to be 7 times the latency of without-ZKRP transactions ($940 ms$ compared to $160 ms$). Separately, \cite{Parsad2024} demonstrate that ZKPs preserves business confidentiality even in transparent environments, but at the expense of computational efficiency.

Our proposal's architecture integrates these privacy and trust requirements while avoiding the overhead of on-chain cryptographic commitments and secondary privacy mechanisms on top of blockchain ledgers. Instead of applying DP or ZKPs directly to the ledger and its data, the architecture utilizes HP private channels to segregate actor data and off-chain SSI and VCs to ensure a smoother more fine grained data exchange between supply chain actors. SSI VCs support ZKPs and selective disclosure, all while remaining verifiable and trusted. Our design isolates the supply chain data acquired by IoT devices and its real-time monitoring and quality authentication using smart contracts from the supply chain data exchange. While the first happens over blockchain using a blockchain-IoT connector, the second happens over SSI systems using blockchain-SSI integration. This allows us to combine the blockchain trust and transparency services with the SSI privacy properties.

\section{Organizational System Model and Requirements}
\label{sec:requirements}

This section provides an organizational overview of the generic supply chain problem, independent of any specific technical implementation. This allows the reader to envisage their own use case.
We first identify the organizational entities that participate in the supply chain and describe their roles. We then derive the system's functional, regulatory, security and privacy requirements, as well as the threat model it must address.

\subsection{Organizational System Model}
\label{subsec:org_model}

\begin{figure*}[t]
  \centering
  \includegraphics[width=0.8\textwidth]{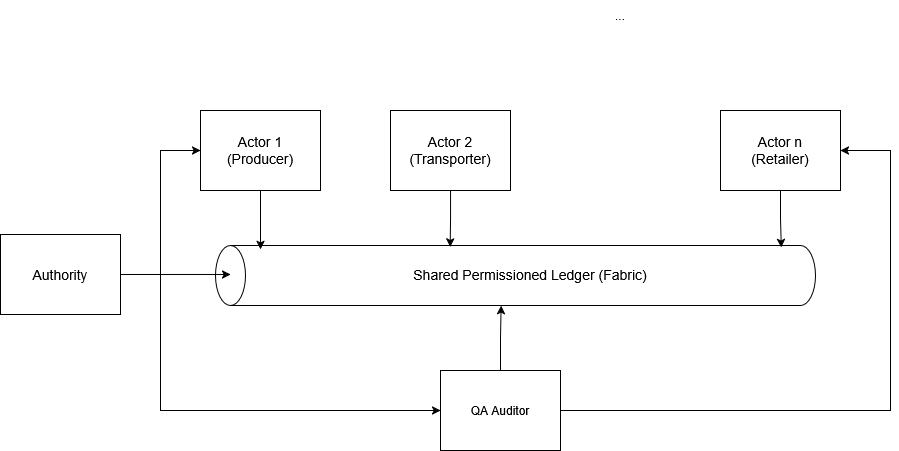}
  \caption{Organizational system model showing the relationship between the supply chain actor chain ($1..n$), the QA auditor, and the authority.}
  \label{fig:actor_chain}
\end{figure*}

We consider a generic multi-stakeholder supply chain composed of a sequence of organizational actors that exchange products and related information under partial trust. At the organizational level, the following entities are distinguished, as shown in Figure~\ref{fig:actor_chain}:

\textbf{Actor 1 (upstream producer).} Actor~1 denotes the upstream organization that creates, harvests, or initially registers a product batch in the supply chain. Depending on the domain, this actor may correspond to a farm, raw-material supplier, or primary producer. Its role is to originate the batch, associate it with its initial production context, and provide the first set of evidences required for provenance, certification, and quality assurance.

\textbf{Actor $i$ (intermediate operator).} Actor~$i$, with $1 < i < n$, denotes any intermediate organization that receives a batch, performs one or more value-adding or logistics operations, and transfers the batch further downstream. Examples include transporters, storage operators, processors, packagers, wholesalers, or distributors. Their role is to maintain continuity of custody, contribute process-specific evidences, and ensure that their operations comply with the applicable contractual and regulatory constraints.

\textbf{Actor $n$ (downstream operator / retailer).} Actor~$n$ denotes the downstream organization that receives the final product before it reaches the end user or consumer-facing market. This actor may correspond to a retailer, final distributor, exporter, or other last commercial operator. Its role is to provide the final chain-of-custody link, support downstream verification of provenance and compliance, and expose the subset of information that must be made available to customers or market-surveillance bodies.

\textbf{QA auditor.} The QA auditor is an external or semi-external organizational entity responsible for verifying compliance with quality, traceability, certification, or process requirements. Its role is not to participate in routine product transformation or logistics, but rather to inspect evidence, verify claims, perform targeted checks on actors or batches, and issue audit conclusions. The QA auditor specifies the quality authentication policies for each product and the requirements for each actor. These policies are used by the QA auditor's quality assurance check smart contracts and applied to each actor's data on the ledger to determine conformity of their services or products with the quality assurance in an automated way. This automation allows the QA auditor to issue certifications and credentials in a streamlined manner. The smart contracts determine conformity, indicating to the QA auditor that they can issue signed certificates and credentials to attest to this conformity after proper check. QA auditors are accountable for their smart contracts.

\textbf{Authority.} The authority denotes a regulatory, supervisory, or certification-recognition body that defines, enforces, or recognizes the rules under which the supply chain operates. Its role is to specify compliance obligations, determine what evidence must be available for regulatory or certification purposes, and, where appropriate, authorize or supervise auditors and certification procedures.

Figure~\ref{fig:actor_chain} summarizes this organizational model as a chain of actors $1..n$ interacting with a QA auditor and an authority. Note that while all actors have their data on the shared permissioned ledger, our architecture enforces a data segregation mechanism as detailed in \ref{subsec:tech_model}.

This model is intentionally technology-agnostic. At this stage, we focus only on organizational roles and expectations. The technical mechanisms supporting these interactions are introduced later in Section~\ref{sec:architecture}.

\subsection{Organizational Functional and Regulatory Requirements}
\label{subsec:org_func_req}

From the above organizational model, the system must satisfy the following functional and regulatory requirements:

\textbf{FRR1 -- Batch origination and identification.} Actor~1 must be able to register the existence of a product batch together with the minimum evidence required to establish its origin, identity, and initial production context.

\textbf{FRR2 -- Chain-of-custody continuity.} Each Actor~$i$ must be able to receive, transform, split, merge, transport, store, or transfer a batch while preserving a verifiable chain of custody across organizational boundaries.

\textbf{FRR3 -- Role-dependent evidence contribution.} Each organizational actor must be able to contribute the evidences that correspond to its own role, such as production information, logistics events, storage conditions, processing steps, certifications, or transfer records.

\textbf{FRR4 -- Targeted quality assurance and audit.} The QA auditor must be able to inspect the evidence associated with a specific batch, time period, actor, or process step in order to verify compliance with predefined rules, standards, or certification procedures. The QA auditor relies on authorized access to an actor's private channel to inspect the data and to deploy quality assurance check smart contracts.

\textbf{FRR5 -- Regulatory supervision.} The authority must be able to define or recognize the rules to be checked, determine the minimum evidence expected from the supply-chain actors, and, where legally justified, obtain access to the information necessary for supervision, investigation, recall management, or enforcement within its mandate. This does not imply unrestricted access to all operational or commercial data.

\textbf{FRR6 -- Selective information disclosure.} The system must support different levels of information disclosure depending on the requesting organizational entity, its mandate, and the purpose of the request. In particular, auditors and authorities may require broader visibility than commercial actors, while consumer-facing or public views should remain limited to non-sensitive information.

\textbf{FRR7 -- Automation of regulatory and certification checks.} The system should support, as a first objective, the automation of the checks explicitly required by regulations, certification schemes, or contractual quality rules. This includes the ability to determine whether required evidences are present, whether constraints have been respected, and whether deviations should trigger flags or alerts. The QA auditor's deployed quality assurance check smart contracts enable this automation.

\textbf{FRR8 -- Evidence-based accountability.} The system must allow actors to justify their compliance by providing verifiable evidence rather than relying only on bilateral trust or unverifiable declarations.

\textbf{FRR9 -- Consumer- or downstream-facing traceability.} Actor~$n$ and other authorized downstream entities should be able to expose a limited but trustworthy view of provenance and compliance information to customers, business partners, or market-surveillance actors.

\textbf{FRR10 -- Heterogeneous organizational participation.} The system must support organizations with different sizes, roles, technical maturity levels, and regulatory obligations, without assuming a single centralized owner of all data.

Overall, the functional and regulatory objective is not unrestricted transparency, but the ability to automate what regulation and quality governance require, while preserving flexibility in how much additional visibility each organization is willing to provide beyond that minimum.

\subsection{Organizational Security and Privacy Requirements}
\label{subsec:org_sec_req}

Beyond functionality, each organizational entity has security and privacy expectations that the system must reconcile with traceability and regulatory compliance.

\textbf{SPR1 -- Regulatory compliance.} The system must guarantee that the minimum information and evidence required by applicable regulations, certification schemes, or supervisory procedures can be made available to the competent auditor or authority.

\textbf{SPR2 -- Business confidentiality.} The system must preserve the confidentiality of commercially sensitive information, including supplier and customer relationships, process parameters, production volumes, pricing logic, sourcing strategies, and operational schedules, unless disclosure is explicitly required. It is particularly important to meet this requirement in order to avoid rejection of the solution by stakeholders.

\textbf{SPR3 -- Controlled transparency.} The system must support a continuum between minimum disclosure and maximum transparency. At the minimum, it must satisfy the disclosure obligations imposed by regulation or audit. At the maximum, an actor may voluntarily expose broader evidence to demonstrate strong compliance or good practices. The architecture must therefore support selectable transparency rather than a single fixed visibility level.

\textbf{SPR4 -- Least-privilege visibility.} Each organizational entity should access only the information necessary for its role and mandate. Commercial actors should not automatically gain access to one another’s full internal data. Auditors and authorities may access broader information, but only within the scope of their function.

\textbf{SPR5 -- Integrity and authenticity of evidence.} Organizations must be protected against forged, altered, or repudiated records. Evidence used for traceability, compliance, or audit must remain attributable to its source.

\textbf{SPR6 -- Accountability of oversight actions.} Actions performed by QA auditors and authorities, such as requests for access, validation decisions, or compliance conclusions, should themselves be traceable and attributable.

\textbf{SPR7 -- Data minimization.} The system should avoid collecting, sharing, or permanently retaining information that is not necessary for operational, contractual, or regulatory purposes.

\textbf{SPR8 -- Purpose limitation.} Information disclosed for one purpose, such as certification, recall management, or food-safety supervision, should not automatically become visible for unrelated commercial or competitive purposes.

\textbf{SPR9 -- Protection against competitive intelligence.} Honest participants should not be forced to reveal enough information for other participants to infer strategic or competitive knowledge beyond what is required by regulation and agreed collaboration.

\textbf{SPR10 -- Support for personal-data protection.} To the extent that records involve personal data, the system must support organizational compliance with data-protection obligations, including data minimization, controlled access, and separation between evidentiary needs and unnecessary exposure.

\subsection{Organizational Threat Model}
\label{subsec:org_threat_model}

At the organizational level, participants are not assumed to be uniformly trustworthy. We distinguish three broad behavioral categories, which may extend to technical components of Section \ref{subsec:tech_threat_model}: 

\textbf{Trusted entities.} A trusted entity is assumed to follow its mandate and not intentionally abuse its position. In a realistic supply-chain setting, it is generally undesirable to assume that all organizations are trusted, because that would trivialize the problem and remove the need for verifiable accountability mechanisms.

\textbf{Honest-but-curious entities.} An honest-but-curious entity follows the prescribed protocol and performs its assigned function correctly, but attempts to infer, retain, or exploit additional sensitive information from the data it can legitimately access.

\textbf{Malicious entities.} A malicious entity may deviate from the protocol in order to conceal non-compliance, misrepresent process conditions, suppress undesirable evidence, forge declarations, gain unauthorized access to data, or exploit observed information for strategic or competitive advantage.\\ 

The following assumptions are realistic for the organizational entities introduced in Section \ref{subsec:org_model}.

\textbf{Actor 1, Actor $i$, and Actor $n$.} Supply-chain actors are modeled as \emph{malicious} with respect to their own business interests. They may execute the following attacks: (i) selectively suppressing, falsifying, or manipulating the timing of submitted evidence, (ii) attempting to access channels, off-chain artifacts, or role-restricted functions without authorization, and (iii) deliberately concealing non-compliance by adjusting their behavior based on the expected level of scrutiny during the audit. Additionally, they may extract information and use it strategically to gain a competitive advantage. This can be done in the following ways: (i) inferring supplier and customer relationships, trading volumes, logistics patterns, and operational schedules; and (ii) reconstructing sourcing strategies, identifying bottlenecks, and determining commercially sensitive process parameters.

\textbf{QA auditor.} The QA auditor is modeled as \emph{honest-but-curious}. Although it is expected to carry out its audit responsibilities according to the established procedures and refrain from falsifying the protocol, it may attempt to extract more information than necessary from the available evidence. This reflects the fact that, for compliance verification purposes, the auditor has broader visibility than ordinary supply-chain actors. However, this creates a risk of excessive information extraction.

Compared to ordinary actors, the QA auditor may have access to richer and more comprehensive information. This information may include detailed process records, sensor histories, certification artefacts, non-public batch trajectories, deviations and incident reports, and evidence spanning multiple actors and time periods. This visibility can lead to the inference of commercially sensitive information, such as supplier and customer relationships, production volumes, operational schedules, process bottlenecks, sourcing strategies, and recurring compliance weaknesses. Even if the auditor follows the audit workflow correctly, they could misuse this additional knowledge, for example, by collecting too much data, retaining it for too long, disclosing it beyond the scope of the audit, or sharing it with unauthorized third parties, such as competitors or other commercial interests. For this reason, an auditor's access must be limited to the batches, actors involved, and time periods covered by the audit engagement. Furthermore, the use of a QA smart contract that applies the QA auditor's policies on the actor's data in the ledger ensures that the inputs and outputs and the conditions of the QA audit are transparent to the actor, the QA auditor and the blockchain peers maintaining the channel. This ensures that the QA process is transparent and respected since it is enforced on the blockchain.

\textbf{Authority.} The authority is considered a \textit{trusted} organizational actor with regard to rule-setting, recognition of schemes, and lawful supervision. However, it is not assumed to control the blockchain infrastructure as a whole, or to have unrestricted visibility of all data. Its primary role is to define, recognize, or publish the rules, compliance schemes, mandates, or supervisory decisions that other participants must follow. This may require write access to specific governance-relevant parts of the system. Read access, by contrast, should only be granted where required by its legal or supervisory mandate. Examples include regulatory oversight, investigation of suspected non-compliance, recall management, and enforcement. This access should remain scoped to the relevant batches, actors, time periods, and evidence categories necessary for the intended purpose. In other words, trust in the authority does not imply universal operational control or unrestricted transparency across all channels and data spaces.

This organizational threat model strikes a balance: it does not assume universal trust, nor does it require the system to tolerate every organizational entity being arbitrarily malicious. Section~\ref{sec:architecture} refines the 
technical implications of these assumptions and introduces the technical components and their trust assumptions.

\section{Reference Architecture}
\label{sec:architecture}

This section introduces the technical reference architecture that implements the organizational requirements of Section~\ref{sec:requirements}. We first define the technical system model and the role of each architectural component. We then state complementary technical security and privacy requirements, followed by the corresponding technical threat model.

\subsection{Technical System Model}
\label{subsec:tech_model}

\begin{figure*}[t]
  \centering
  \includegraphics[width=1\textwidth]{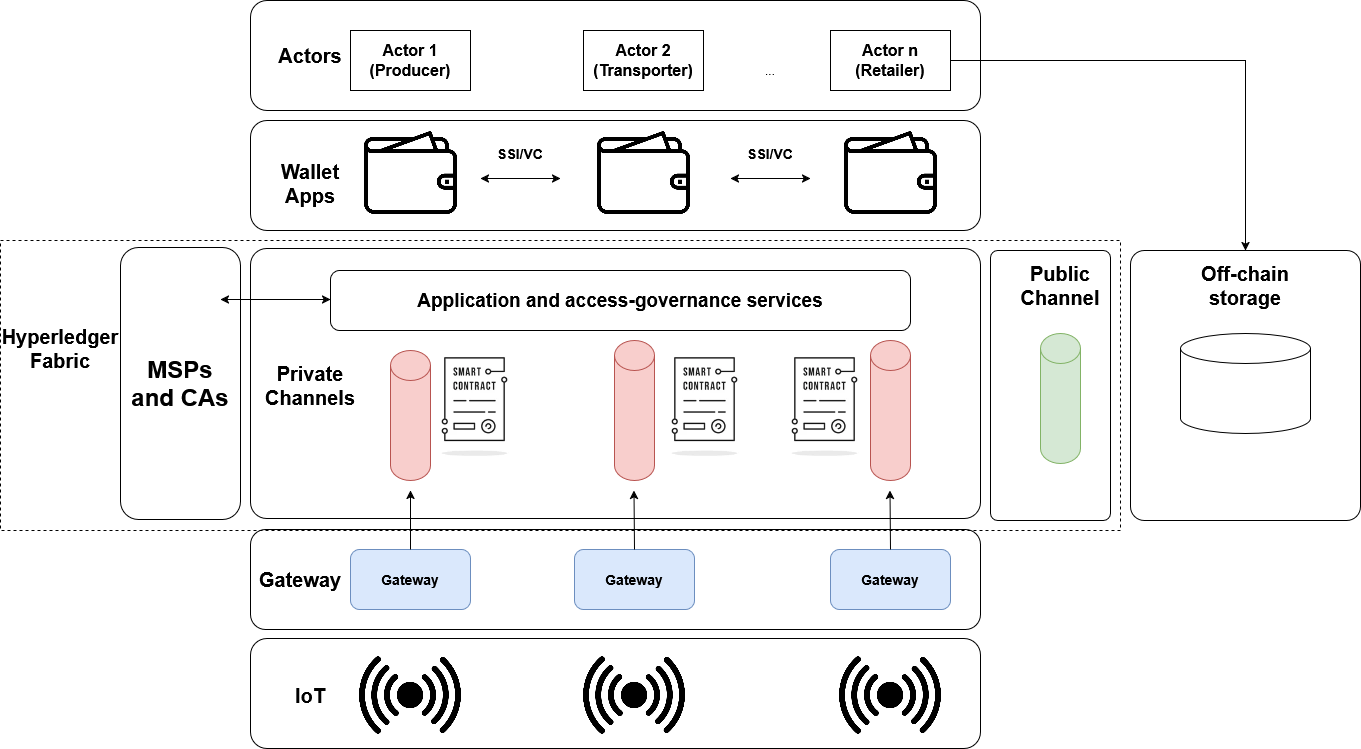}
  \caption{Technical system model of the reference architecture, showing organizational entities, wallets, IoT devices, gateways, Hyperledger Fabric peers and MSPs, public and private channels, smart contracts, and off-chain storage. Private channels provide restricted read/write visibility for subsets of organizations, while off-chain storage retains sensitive artefacts under controlled access.}
  \label{fig:architecture}
\end{figure*}

Figure~\ref{fig:architecture} presents the technical system model of the proposed architecture. This architecture maps each organizational entity introduced in Section~\ref{sec:requirements} to one or more technical components that support identity management, transaction submission, evidence storage, and controlled access to data. Some of these components, such as wallets, IoT devices, and gateways, are operated by specific entities, while others, such as channels and their associated policies, comprise shared governance structures spanning multiple organizations.

\textbf{Wallets of organizational entities.} Each organizational entity (Actor~1, Actor~$i$, Actor~$n$, QA auditor, authority) is associated with one or more digital wallets or client-side identity stores. These wallets hold the credentials used to authenticate the entity toward applications, gateways, and ledger-facing services. In the SSI/VC-enhanced variant, wallets also hold verifiable credentials and support the generation of verifiable presentations (cf. Section~\ref{subsec:ssi_vc}).

\textbf{IoT devices.} IoT devices capture measurements related to product condition, environment, location, or process execution. Several deployment models are possible, depending on the business and operational setting. For example, devices may be: (i) owned and operated directly by the supply-chain actor that uses them; (ii) leased, maintained, or remotely managed by a specialized IoT provider on behalf of a supply-chain actor or platform operator; (iii) owned by a platform operator and deployed across participating organizations; or (iv) jointly managed in shared logistics or warehousing scenarios. These distinctions are security-relevant, because device ownership and operational control influence key provisioning, software updates, physical protection, attribution of measurements, and the extent to which other parties must trust the device operator. The architecture therefore does not assume a single universal deployment model, but requires that the responsible entity and the provenance of the measurements be identifiable.

\textbf{Gateways.} Gateways collect data from IoT devices, perform protocol translation and preliminary validation, associate measurements with the relevant batches or operational context, and submit the resulting evidence to the appropriate technical backend. Responsibility for gateway management may also vary across deployment scenarios. A gateway may be operated by the same supply-chain actor that operates the devices, by a platform operator coordinating the overall system, or by a specialized IoT service provider that manages the ingestion infrastructure. Because gateways are located at the boundary between devices and ledger-facing services, the way they are managed has direct security implications. This affects admission control, credential handling, data filtering and aggregation, replay protection, and the ability to suppress, alter, or fabricate measurements before submission. For this reason, the gateway operation should be attributable to a specific organizational entity.

\textbf{Hyperledger Fabric ledger.} The permissioned ledger provides shared state management, transaction ordering, endorsement, and validation across organizations. It is used as the core coordination layer for tamper-evident recording of provenance events, compliance-relevant states, and integrity anchors for off-chain data.

\textbf{Public and private channels.} The Fabric network is partitioned into channels with different visibility scopes. A public or consortium-wide channel contains integrity-critical and non-sensitive information intended to support broad traceability. In addition, one or more private channels are created for subsets of organizations that must exchange more sensitive operational information. A channel is therefore not simply ``owned'' by a single organization in the application sense; rather, it is jointly governed by the organizations that are members of that channel, according to its configuration, endorsement policies, and access-control rules. Channel membership determines which organizations may host peers that replicate the channel ledger and execute or validate its smart contracts, while application-layer policies may further restrict which functions and artefacts are visible to particular roles within those organizations.
Section \ref{subsec:data_placement} provides a detailed overview of channel organization and data placement.

\textbf{MSP.} Each participating organization is represented in the Fabric network through its own Membership Service Provider (MSP) configuration. The MSP defines the trust anchors and certificate-validation rules by which identities are recognized as belonging to that organization, including peer, client, and administrative identities. In this sense, the organization is not merely a client of an external MSP; rather, its MSP constitutes the organizational identity domain through which it participates in the permissioned network. Access to a private channel is then determined collectively by the channel configuration and policies, which reference the MSPs of the organizations admitted to that channel. The MSP therefore supports channel participation and policy enforcement by defining which identities are valid organizational members, while the channel itself defines the shared governance space in which subsets of organizations collaborate.

\textbf{Off-chain storage.} Larger, more sensitive, or less frequently accessed artefacts are stored outside the ledger. This includes detailed sensor logs, documents, certificates, reports, and other payloads for which direct on-chain storage would be inefficient or would expose excessive detail. The ledger stores only the metadata, references, or cryptographic commitments required to bind these artefacts to the shared process.

\textbf{Smart contracts.} Smart contracts implement the shared logic of the system, including batch lifecycle management, handover recording, compliance flagging, and verification of access preconditions for protected operations. They define the technical state transitions that correspond to the organizational operations of Section~\ref{sec:requirements}. There are two types of smart contract specified in our proposal: QA smart contracts and real-time monitoring contracts. QA contracts are deployed by the QA auditor alongside the QA policy on the actor's private channel when the QA process is executed. Real-time monitoring contracts are deployed by actors on their own channels, and they use the bounds defined in the QA policy to create events and notify actors about the status of compliance of their product/service using IoT data.

\textbf{Application and access-governance services.} On top of the ledger, application-layer services provide business interfaces for the different organizational entities. These services are also the enforcement point for SSI/VC-based access governance, fine-grained authorization, and controlled release of off-chain artefacts.

\subsection{Specific Focus on Hyperledger Channel Organization and Data Placement}
\label{subsec:data_placement}

The central design principle is that not all information is stored in the same technical space or with the same level of visibility. Instead, the architecture combines public visibility for situations requiring broad traceability, restricted channel visibility for situations requiring operational confidentiality, and off-chain storage for situations requiring tighter control of detailed artefacts.

\textbf{Private-channel organization and access semantics}
Because private channels are the core of the proposal, their role must be made explicit.

A private channel is controlled by the subset of organizations that are members of that channel. Membership determines which organizations can host peers that replicate the channel ledger and execute or validate its smart contracts. Within a given private channel, write access is governed by the endorsement and submission policies of the corresponding chaincode, while read access is governed by channel membership and application-layer access policies.

Different private channels can serve different purposes. For example, one channel may support bilateral or local operational exchange between adjacent actors in the value chain, while another may support an audit-oriented view shared with a QA auditor for a specific subset of compliance-relevant data. Within such channels, different data zones can be distinguished conceptually: (i) operational event data required by the participating actors, (ii) compliance-relevant summaries or flags, and (iii) references or commitments to off-chain detailed artefacts. This distinction clarifies that not every channel member must necessarily receive the same level of semantic detail at the application layer, even if the channel provides shared ledger replication among its members.

\textbf{Data placement principles}
Three broad classes of data are handled by the architecture: (1) public or broadly shareable provenance anchors, (2) restricted operational and compliance data, and (3) detailed or sensitive artefacts kept off-chain. This separation is essential to support both regulatory compliance and business confidentiality.

\subsection{Complementary Technical Security and Privacy Requirements}
\label{subsec:tech_sec_req}

The technical architecture must meet the following technical requirements in order to satisfy the organizational security and privacy requirements set forth in Section~\ref{subsec:org_sec_req}. Please refer to Section~\ref{sec:background} for a better technical concept understanding.

\textbf{TR1 -- Authentication of technical components.} The architecture must authenticate users, organizations, devices, gateways, and services before they can submit, relay, or access protected operations. This is done using MSP X.509 certificates for ledger authorization and using DIDs and DID-Auth to authenticate actors and components off-ledger.

\textbf{TR2 -- Fine-grained authorization.} The architecture must support authorization decisions based on organizational role, technical identity, channel membership, and, where applicable, verifiable credentials and mandate attributes. This is supported by the SSI layer's SSI credentials (VCs), as well as smart contracts on the ledger combined with MSP capabilities and certificates.

\textbf{TR3 -- Channel-based confidentiality.} The architecture must restrict the visibility of ledger data according to channel membership and associated access policies. This is implemented by HF private channels and MSP authorization.

\textbf{TR4 -- Controlled off-chain access.} Access to off-chain artefacts must be protected independently of simple possession of a ledger reference, and must remain subject to explicit access-control checks. VCs on the SSI layer are used to control off-chain access, where the data provider generates an access VC for the data consumer. Furthermore, the data provider can anchor access decisions and related details on their private channel to ensure accountability through the ledger.

\textbf{TR5 -- Integrity binding between on-chain and off-chain data.} Any off-chain artefact referenced by the system must be cryptographically bound to the ledger-visible evidence that points to it. Sensitive or bulky data should remain off-chain unless direct on-chain storage is strictly necessary. Hash commitments of off-chain actor data are published on the private channel of the actor.

\textbf{TR6 -- Attributability of submissions.} All transactions, device-originated measurements, gateway submissions, and administrative actions must be attributable to the technical entity (and therefore the organizational entity) that performed them. Since all actors and components are identified using DIDs, all entities must sign their data, which can be verified using the public key associated with the DID. Furthermore, all ledger actions are authenticated using HF certificates. 

\textbf{TR7 -- Tamper evidence.} The architecture must make unauthorized modification or deletion of recorded evidence detectable. All blockchain records are tamper proof. Moreover, from TR5 the off-chain data integrity is guaranteed through ledger commitments.

\textbf{TR8 -- Segregation of duties.} No single technical component should be assumed to unilaterally control all identity, evidence ingestion, storage, and validation functions.

\textbf{TR9 -- Support for selective information disclosure.} The architecture should only release the minimum number of attributes or artefacts required for a given request. The VC schema of the SSI layer supports selective information disclosure.

\textbf{TR10 -- Auditability of access decisions.} Technical actions related to authorization, disclosure and compliance should be logged in a way that can be attributed to a specific person and subject to audit. Smart contracts are used to log events and all access decisions on the ledger.

\subsection{Complementary Technical Threat Model}
\label{subsec:tech_threat_model}

We now refine the threat model at the level of technical components. As in Section~\ref{subsec:org_threat_model}, components may be trusted, honest but curious, or malicious, depending on the assumptions made.


\textbf{Wallets.} Wallets may behave \emph{maliciously} either by being compromised or being under the control of malicious organizational entities. A malicious wallet can attempt to submit false credentials, trying to corrupt the system with fake data.
Therefore, the architecture must limit the consequences of a malicious wallet by implementing scoped credentials and authorization checks.
We suppose it never leaks the data of its organizational actor.

\textbf{IoT devices.} IoT devices may behave \textit{maliciously} if compromised, cloned, tampered with, or intentionally misconfigured, and may also be unreliable due to operational faults. The behavior of these devices depends on who owns, provisions, maintains, and physically controls them. For example, a device operated by a specialized IoT provider raises different issues than a device operated directly by a supply-chain actor or platform operator. Therefore, their outputs should not be accepted blindly without attribution, validation, and contextual checks.
We assume that IoT device data cannot be exfiltrated by other actors.

\textbf{Gateways.} Gateways may behave \textit{maliciously}. They may alter, suppress, replay, or fabricate measurements before submission. 
The risk profile depends partly on which organizational entity operates the gateway, since actor-, platform-, and third-party-managed gateways imply different trust and accountability assumptions. Gateways therefore require attributable identities, controlled credentials, and auditable submission behavior.
We suppose that Gateway aggregated data cannot be exfiltrated by external actors.

\textbf{Hyperledger Fabric peers and ordering services.} Due to the consortium based management, the ledger infrastructure is modeled as \textit{trusted}, since it is operated by an honest majority via a fault-tolerant consensus protocol.

\textbf{Smart contracts.} Smart contracts are \textit{trusted} only to the extent that their code is correct, consistently deployed, and can be verified by peers.

\textbf{Off-chain storage.} Off-chain storage providers are \textit{malicious}. 
They may attempt to delete or alter content. 

\textbf{Application and access-governance services.} These services are powerful enforcement points and are modeled as \textit{trusted}. MSPs and Certificate Authorities (CAs) are operated by different organizations, each of which provides access to its relevant clients. An actor only needs to trust their associated CAs or MSP.

\section{Security and Privacy Analysis}
\label{sec:security}

This section evaluates the security and privacy properties of the proposed generic architecture. The analysis is conducted in light of the organizational requirements and threat model introduced in  Section~\ref{sec:requirements}, and the technical system model, technical requirements, and technical threat model introduced in Section~\ref{sec:architecture}. For simplicity, SPRs are grouped together under broader categories. Table \ref{tab:req_sat} provides a direct mapping between the security and privacy requirements (SPRs) defined and the functional, regulatory and technical requirements of our architecture that satisfy them.

\subsection{Evidence Falsification and Competitive Intelligence (SPR1, SPR2, SPR5, SPR9)} 
Actors (1...n) in the supply chain are modeled as malicious when it comes to their business interests. Our proposed architecture mitigates their attempts to suppress, repudiate or falsify evidence of non-compliance by compromising their wallet, gateways, IoT devices or off-chain storage components. This is achieved through the use of immutable cryptographic binding between on-chain and off-chain data (TR5), as well as tamper evidence (TR7), which uses blockchain and data hashes. Attributable submissions and evidence (TR6) ensure that all supply chain data and events are non-repudiable, as they are recorded on-chain too. The segregation of duties (TR8) also limits actors' capacities to suppress or falsify data, as they do not have unilateral control over it.
Furthermore, certain functional and regulatory requirements, such as chain-of-custody continuity (FRR2) and evidence-based accountability (FRR8), ensure that data cannot be repudiated or altered.
In the same context, malicious actors may attempt to extract data exploit ledger data to gain competitive intelligence (SPR9) by inferring relationships, and production volumes, among other things, about their supply chain competitors. We mitigate this risk through data segregation on HF, where each actor's data is stored in a separate private channel (TR3) created specifically for them to protect their business confidentiality (SPR2).

\subsection{Controlled Governance and Authorization (SPR3, SPR4, SPR6, SPR8)}
One of the supply chain problems that our architecture addresses is placing full trust in governance and audit entities. In our model, authorities and QA auditors are assumed to be honest-but-curious, so even though they require broader visibility, we acknowledge the risk of them extracting information excessively. To restrict this, we use SSI VC credentials for fine-grained authorizations (TR2) and purpose limitation (SPR8), while also immutably logging all ledger access decisions (TR10) satisfying SPR3 and SPR4. This enables actors to select how to authorize the QA auditors and the authorities to access their private channels, while providing an irrefutable record of access and its purpose (SPR6). This mitigates mandate abuse.

\subsection{Secure Data Ingestion and Minimization (SPR7, SPR10)}
The boundary between the real world and the architecture is formed by IoT devices and gateways. Securing their data is crucial for the architecture to function correctly. We limit the vulnerabilities of compromised or misconfigured components through authentication (TR1) and attributability (TR6). Furthermore, we provide data minimization (SPR7) and personal data protection (SPR10) by using off-chain storage for sensitive or bulky data and ensuring their integrity via on-chain anchoring (TR5).

\begin{table*}[htbp]
\centering
\caption{Mapping between security and privacy requirements, and the technical requirements needed to satisfy them}
\label{tab:req_sat}
\renewcommand{\arraystretch}{1.3}
\begin{tabular}{p{0.22\linewidth}p{0.15\linewidth}p{0.15\linewidth}p{0.38\linewidth}}
\hline
\textbf{Security/Privacy Requirement (SPR)} & \textbf{Functional and Regulatory Requirements (FRR)} & \textbf{Technical Requirements (TR)} & \textbf{Reasons why requirements are met} \\ \hline
SPR1: Regulatory compliance & FRR1, FRR5, FRR7 & TR5, TR7 & Minimum viable compliance evidence is anchored on-chain with immutable cryptographic binding. \\ 
SPR2: Business confidentiality & FRR6 & TR3, TR4, TR5 & Proprietary operational data is isolated within private channels and restricted off-chain storage. \\ 
SPR3: Controlled transparency & FRR3, FRR6 & TR2, TR3, TR9 & Visibility operates on a selectable continuum governed by Verifiable Credentials (VCs). \\ 
SPR4: Least-privilege visibility & FRR4 & TR2, TR3, TR4 & Actor and auditor access is technically bounded by explicit role and channel membership constraints. \\ 
SPR5: Integrity and authenticity & FRR2, FRR8 & TR1, TR5, TR6, TR7 & Device telemetry and actor submissions are authenticated, attributable, and tamper-evident. \\ 
SPR6: Accountability of oversight & FRR4, FRR5 & TR6, TR10 & Governance actions (e.g., QA auditor requests) are immutably logged to prevent mandate abuse. \\ 
SPR7: Data minimization & FRR6 & TR5, TR9 & Ledger payload is minimized; sensitive artifacts remain off-chain unless explicitly required. \\ 
SPR8: Purpose limitation & FRR6 & TR2, TR4, TR9 & SSI selective information disclosure prevents data authorized for compliance from being reused for intelligence. \\
SPR9: Competitive intelligence protection & FRR6, FRR10 & TR3, TR5, TR9 & Channel partitioning and off-chain data placement eliminate global ledger analysis vectors. \\ 
SPR10: Personal-data protection & FRR6 & TR5, TR9 & Identity abstraction and off-chain storage align with data minimization and controller mandates. \\ \hline
\end{tabular}
\end{table*}

\section{MoreMedDiet Proof-of-Concept}
\label{sec:implementation}

The MoreMedDiet proof-of-concept (PoC) instantiates the
reference architecture in the context of a sustainable
Mediterranean agri-food supply chain. The PoC demonstrates how IoT-based evidence collection,
Hyperledger Fabric private channels, a blockchain mediation
layer, and SSI/VC-based access governance can be combined
to enable controlled transparency between partially trusted
supply-chain actors.

\subsection{MoreMedDiet-Specific Instantiation}

In the MoreMedDiet scenario, the generic actors introduced
in Section III are mapped to a farm-to-fork agri-food chain, as reported in Table ~\ref{tab:mmd-mapping}. 
Primary producers, transporters, processors, distributors, and
retailers correspond to the product-flow actors. Each actor
maintains evidence about the product attributes and events
that fall under its operational responsibility.


The specificity of the MoreMedDiet PoC is its combination
of product provenance, food-quality evidence, transport and
storage-condition monitoring, and selective inter-actor
disclosure. The architecture does not assume that all actors
share their full operational data with the entire consortium.
Instead, each actor records their own evidence in a private
ledger context and discloses selected claims or proofs when another actor or auditor requests them for legitimate reasons. This enables the platform to support traceability and
quality assurance without exposing commercial
information that is not necessary, such as detailed routes, volumes, supplier
relationships, or internal process parameters.


\begin{table}[t]
\centering
\caption{Instantiation of the reference architecture concepts in the MoreMedDiet PoC}
\label{tab:mmd-mapping}
\begin{tabular}{p{0.34\columnwidth}p{0.54\columnwidth}}
\hline
\textbf{Reference concept} & \textbf{MoreMedDiet PoC instantiation} \\
\hline
Supply-chain actor &
Producer, transporter, processor, distributor, or retailer participating in the agri-food chain. \\
Actor-specific evidence space &
Private Hyperledger Fabric channel associated with the actor and its product-related evidence. \\
IoT devices &
Sensors collecting measurements related to product or environmental conditions. \\
IoT gateway &
Component that aggregates readings from IoT devices and forwards them to the blockchain mediation layer. \\
Blockchain Agent &
Mediator component that receives data from gateways or applications and invokes the relevant smart contract functions. \\
Smart contracts &
Logic for inserting readings, updating product attributes, recording evidence references, and supporting QA inspection. \\
SSI/VC layer &
Mechanism used by actors to request, issue, present, and verify claims derived from private-channel evidence. \\
QA access &
Scoped access allowing a QA actor to inspect the relevant private-channel history of product attribute changes. \\
\hline
\end{tabular}
\end{table}

\subsection{Prototype Platform and Data Flow}

\begin{figure*}
    \centering
    \includegraphics[width=0.9\linewidth]{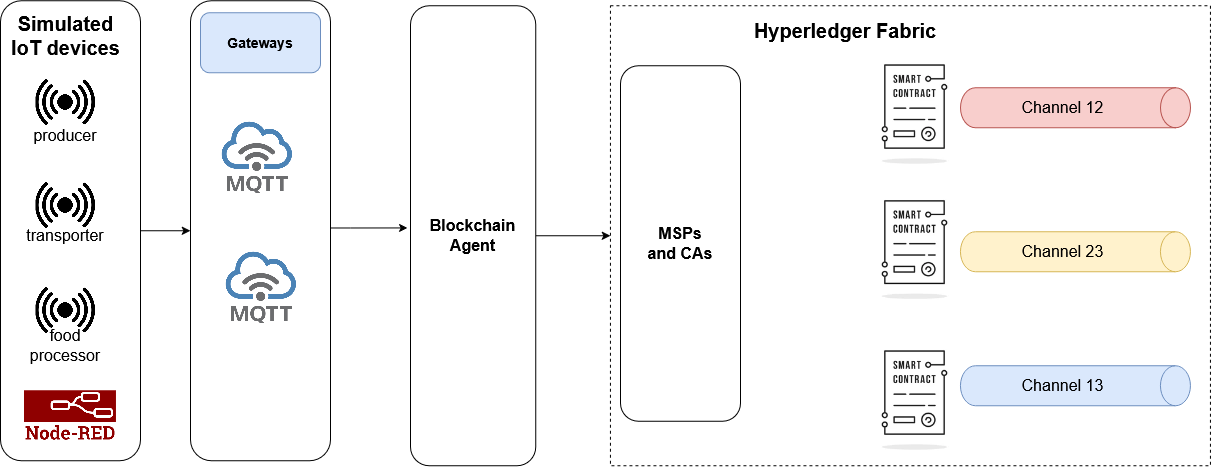}
    \caption{The MoreMedDiet Proof-of-concept.}
    \label{fig:pocsim}
\end{figure*}

Figure \ref{fig:pocsim} provides an overview of the implemented PoC, in which Node-Red \cite{Nodered} is used to simulate data from IoT devices and MQTT \cite{MQTT} brokers act as IoT gateways, aggregating and publishing data to the private channels of each actor via the Blockchain Agent. The Blockchain Agent is deployed on a Ubuntu 24.04.2 Docker container. Each actor has a private channel assigned to them, which acts as an evidence space, accessible through MSP permissions and certificates. The deployed blockchain comprises three peer nodes and a single orderer, forming three private channels (12, 13 and 23) between peer 1 and peer 2, peer 1 and peer 3, and peer 2 and peer 3, respectively. Each simulated actor is assigned a private channel. We use Eclipse Mosquitto 2.0 deployed on Docker containers to simulate the IoT gateways. We developed an Android mobile application as an SSI wallet to enable actors to exchange credentials and interact with the blockchain. It was deployed on an Android 14 emulator with an ARM64 processor and 2048 MB of RAM. Apart from the SSI wallet mobile application, all of components are deployed as Docker containers on the same server. Network latencies are omitted since all components operate on the same local area network (LAN).


The Blockchain Agent acts as a mediation component between the application and IoT layers, and the Hyperledger Fabric network. It validates incoming requests at the platform boundary, selects the relevant smart contract operation, and invokes the transaction that inserts corresponding readings or evidence into the private channel associated with the relevant actor.

IoT devices do not submit transactions directly to the ledger. Instead, IoT gateways aggregate readings from one or more devices and forward the resulting data to a Blockchain Agent. This design separates constrained or heterogeneous IoT devices from ledger-specific logic. The gateway  collects and structures device data, while the Blockchain Agent handles Fabric-specific interactions, including transaction submission and smart contract invocation.

\begin{figure*}
    \centering
    \includegraphics[width=0.9\linewidth]{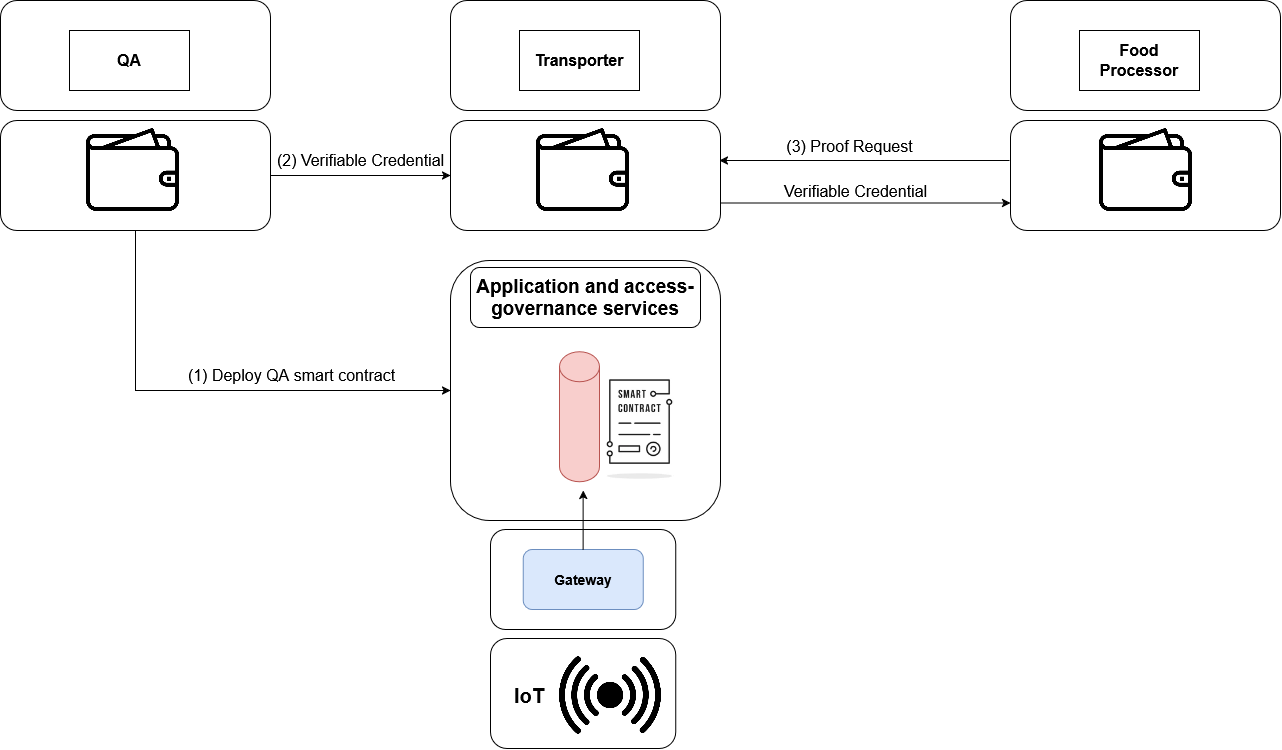}
    \caption{MoreMedDiet data flow illustration when a food processor requests proof from a transporter that they have met the requirements.}
    \label{fig:scen1}
\end{figure*}

The mobile application provides a user interface for selective information disclosure between actors. For example, a food processor can request proof of transport conditions from a transporter. Figure \ref{fig:scen1} illustrates  this process: thanks to the ledger data collected by IoT devices and a QA smart contract deployed by the QA with authorization from the food processor (step 1), the QA can issue certifications and credentials that prove that the transporter has met the requirements (step 2). Before accepting a shipment, the food processor requests such proofs (step 3) that the transporter presents (step 4).

Table~\ref{tab:mmd-workflows} summarizes the main
workflows validated in the PoC.

\begin{table}[t]
\centering
\caption{Main workflows validated in the MoreMedDiet PoC}
\label{tab:mmd-workflows}
\begin{tabular}{p{0.28\columnwidth}p{0.46\columnwidth}p{0.16\columnwidth}}
\hline
\textbf{Workflow} & \textbf{Description} & \textbf{Status} \\
\hline
IoT reading collection &
IoT devices generate readings related to product or environmental conditions. &
Implemented \\
Gateway aggregation &
The IoT gateway aggregates readings and prepares them for submission to the platform. &
Implemented \\
Blockchain mediation &
The Blockchain Agent receives gateway or application data and invokes the relevant Fabric smart contract. &
Implemented \\
Private-channel insertion &
Readings and product-attribute updates are inserted into the private channel associated with the responsible actor. &
Implemented \\
Inter-actor VC request &
One actor requests a claim or proof from another actor, for example regarding transport conditions. &
Implemented \\
VC-based disclosure &
The responding actor provides a verifiable proof derived from private-channel evidence. &
Prototype \\
QA inspection &
A QA actor obtains scoped access to inspect the history of product attribute changes in an actor's private channel. &
Prototype \\
Credential revocation &
Revocation of previously issued credentials or access rights. &
Under consideration \\
\hline
\end{tabular}
\end{table}

\subsection{Validation and Performance Characterization}

Based on the methodology detailed in Appendix \ref{app:benchmarking-methodology}, we obtained the results described in Table~\ref{tab:mmd-characterization}. The table presents the mean, median and 95th percentile (p95) of key operations evaluated on the PoC.

\begin{table}[t]
\centering
\caption{Prototype evaluation results}
\label{tab:mmd-characterization}
\begin{tabular}{p{0.22\columnwidth}p{0.32\columnwidth}p{0.36\columnwidth}}
\hline
\textbf{Operation} & \textbf{Value / metric} & \textbf{Purpose} \\
\hline

Gateway-to-agent submission &
Mean 166 ms; median 165 ms; p95 178 ms &
IoT ingestion path. \\

Agent-to-ledger transaction &
Mean 50 ms; median 53 ms; p95 57 ms &
Mediated Fabric write path. \\

Private-channel query &
Mean 53 ms; median 56 ms; p95 61 ms &
Evidence retrieval path. \\

VC proof generation &
Mean 31 ms; median 31 ms; p95 34 ms &
Disclosure preparation overhead. \\

VC proof verification &
Mean 52 ms; median 30 ms; p95 32 ms &
Access-governance verification overhead. \\
\hline
\end{tabular}
\end{table}
As the table shows, the average latency for interactions between the agent and the blockchain is $50 ms$, while the latency for interactions between the gateway to the agent is around $160 ms$. Therefore, writing aggregated IoT data from the device to the blockchain only takes an average of around $210 ms$. This latency is viable, since while IoT devices monitor and collect data at all times, the aggregated data is only written a few times a day in batches \cite{Gug}. Standard HF deployments transaction latencies are in the order of hundreds of milliseconds \cite{Istivan}, making our performance metrics viable for multi-actor supply chains and for aggregated batch data.

Furthermore, retrieving data from private channels only takes an average of just $50 ms$, making it highly efficient for retrieving evidence in order to generate credentials \cite{Ali}. Generating a credential takes $30 ms$, and verifying it takes $50 ms$. Therefore, an actor would need to spend around $130 ms$ in total to retrieve proofs from the blockchain, generate a credential and present it to another actor for verification. 
These figures should be compared to the typical credential transfer times observed in the supply chain, which generally range from a few hours to a few days \cite{wallmart}. They should also be compared to the time of 2.2 seconds achieved for certifying the origin of mangoes in Walmart’s supply chain \cite{wallmart}. It is also recognized that latencies approaching $100 ms$ are perceived as instantaneous by users in Human-computer interfaces (HCI) and mobile environments \cite{responselatency}. 
Therefore, we can conclude that a delay of $130 ms$ is near real-time for the supply chain sector, and that the model is highly likely to be applicable to real-world supply chain environments.




\section{Conclusions and Future Work}
\label{sec:conclusion}


Integrating IoT for real-time and continuous monitoring of supply chain events with a consortium blockchain infrastructure helps automate and accelerate logging, certification and exchange of trusted data between supply chain actors. Using data segregation through private channels and fine-grained authorization through MSP certificates and verifiable credentials, we are able to satisfy legal and privacy requirements and most importantly controlled transparency that protects the business interest of actors against competitive intelligence. Furthermore, through the use of off-chain storage and delegation of cryptographic proofing burdens and authentic data exchange to the SSI layer, we avoid the limitations and burdens of on-chain proofs. The proposed architecture and its instantiation for an agri-food supply chain project combined with a Proof-of-Concept, its evaluation results and the security and privacy analysis, prove feasibility, potential applicability of the model to real-world supply chain environments and reliability of our proposal.
Nevertheless, there is still work to be done. The revocation of credentials remains an open issue, since many different methods exist for revoking VC credentials, e.g. cryptographic accumulators \cite{accum}, W3C Bitstring Status Lists \cite{BSL} and more privacy-oriented solutions like CRSet \cite{CRSet}. The performance and privacy implications of each method still need to be explored in the context of our proposal.
Furthermore, integrating zero-knowledge proofs into credentials would better protect the business activity of stakeholders by ensuring that only the necessary data and proof is delivered to different actors, e.g. confirming that the temperature was below the required threshold without revealing the true temperature. Furthermore, neither the prototype nor the evaluation data take into account network latencies or large-scale use cases. High-stress testing across an expanded network is therefore required to evaluate real-time scenarios.


\appendix

\section{Benchmarking Methodology}
\label{app:benchmarking-methodology}

This appendix describes the methodology used to obtain the
performance-characterization values reported for the
MoreMedDiet proof-of-concept. The measurements are intended
to characterize the experimental prototype deployment rather
than to provide general performance benchmarks for all
possible deployments of the reference architecture. In
particular, the results depend on the local network, container
configuration, Hyperledger Fabric setup, device profile, and
software versions used in the experiment.

\subsection{Experimental Setting}

The benchmarking was performed on the MoreMedDiet
experimental deployment. In this deployment, IoT gateways
publish sensor events through MQTT, and the Blockchain
Agent acts as an intermediary between the MQTT-based
ingestion layer and the Hyperledger Fabric network. The
Blockchain Agent receives or subscribes to incoming sensor
events, processes them, and invokes the appropriate smart
contract operations to insert or retrieve data from the relevant
Fabric channel.

The measured ledger operations were executed against the
prototype Fabric network used by the PoC. The network
included two organizations, two peers, and a Raft ordering
service. The latency measurements therefore reflect the
behavior of this local experimental setup and should not be
interpreted as deployment-independent properties of the
architecture.

\subsection{Gateway-to-Agent Submission Latency}

Gateway-to-Agent latency characterizes the IoT ingestion
path between the MQTT gateway layer and the Blockchain
Agent. During the benchmark, timestamped MQTT messages
were published to the MQTT broker used by the Blockchain
Agent. Each message contained a unique identifier, a
publication timestamp, and a synthetic sensor value.

For each message, latency was computed as the difference
between the timestamp at publication and the timestamp
observed when the message was received during the
measurement campaign. This metric therefore captures the
message delivery delay between the MQTT publication point
and the receiving side of the Blockchain Agent pipeline. It
does not include subsequent Fabric transaction endorsement,
ordering, or commitment.

A total of 1000 MQTT messages were used for this
measurement. The reported values correspond to the mean,
median, and 95th percentile of the collected samples.

\subsection{Agent-to-Ledger Transaction Latency}

Agent-to-Ledger latency characterizes the mediated Fabric
write path. The benchmark repeatedly invoked a smart
contract function through the Hyperledger Fabric command
line interface. For each iteration, the start timestamp was
recorded immediately before submitting the transaction
invocation, and the end timestamp was recorded after the
invocation completed.

This metric captures the elapsed time for submitting a
transaction to the Fabric network in the prototype
environment, including the interaction with the ordering and
peer infrastructure required for the invocation to complete.
It does not include MQTT publication latency or mobile
application processing.

A total of 100 transaction invocations were executed for
this measurement. The reported values correspond to the
mean, median, and 95th percentile of the collected samples.

\subsection{Private-Channel Query Latency}

Private-channel query latency characterizes the evidence
retrieval path used by the Blockchain Agent when reading
data from the Fabric ledger. The benchmark repeatedly
executed ledger query operations against the deployed smart
contract. For each query, the start timestamp was recorded
immediately before the query request and the end timestamp
after the query completed.

This metric captures the response time of the query path in
the prototype deployment. Unlike transaction invocations,
queries do not require ordering in Fabric; therefore, this
measurement primarily reflects peer-side query processing,
chaincode execution for the query, and local communication
overhead in the testbed.

A total of 100 query operations were executed for this
measurement. The reported values correspond to the mean,
median, and 95th percentile of the collected samples.

\subsection{Application-Side VC Proof Measurements}

VC proof generation and verification were measured directly
inside the running mobile wallet application. The benchmark
was executed automatically 30 seconds after wallet
initialisation, allowing background setup tasks to complete
before timing began.

To isolate the cryptographic cost of proof generation and
verification, the benchmark constructed a complete AnonCreds
credential stack in memory inside the application process. No
network communication, ledger access, mediator call, or
external service call was performed during the measured
operations. This avoids mixing the cryptographic cost of the
SSI/VC workflow with network latency or backend
availability.

Before measurement, a synthetic credential was prepared in
memory. The credential schema contained three attributes:
the access requester, the access granter, and the resource
owner. A credential definition was derived from this schema
using the Camenisch-Lysyanskaya (CL) signature scheme \cite{CL} (without revocation support). A
prover link secret was generated, a credential request was
formed, and a credential containing concrete attribute values
was issued and stored locally in the application. This
credential-preparation phase was treated as setup and was
excluded from the reported timing values.

Five proof-exchange iterations were then executed. Each
iteration measured two operations independently. The first was
VC proof generation, defined as the time required for the
prover to construct a zero-knowledge presentation proving
possession of the credential attributes required by the
presentation request. The second was VC proof verification,
defined as the time required for the verifier to validate the
presentation against the presentation request, schema, and
credential definition.

Each iteration used the same locally stored credential but a
fresh presentation request with a unique nonce. This reflects
the intended use of independent proof exchanges while
keeping the credential material constant across iterations. The
cryptographic operations were performed using the
Hyperledger AnonCreds Rust library through its Android
bindings. The measurements therefore reflect the cost of the
actual AnonCreds cryptographic implementation used by the
mobile application, rather than a simplified simulation.

The mobile benchmark was executed on an Android Virtual
Device using the arm64-v8a architecture. The reported values
were computed from five samples per operation and include
the mean, median, and 95th percentile in milliseconds.

\subsection{Reported Statistics and Limitations}


The results should be interpreted with the following
limitations. First, the measurements were collected in a local
experimental deployment and are sensitive to the host
machine, container runtime, network setting, and Fabric
configuration. Second, the application-side VC measurements
intentionally exclude network communication and ledger
access, and therefore represent cryptographic proof-processing
costs only. Third, the benchmark does not constitute a
large-scale stress test of the architecture; rather, it provides a
reproducible characterization of the main execution paths
validated by the PoC.

\section*{Acknowledgements}
We would like to express our sincere thanks to the members of the MoreMedDiet consortium, and the Agence Nationale de la Recherche (ANR) who funded the MoreMedDiet project.

\section*{Funding}
This research was funded in whole or in part by the Agence Nationale de la Recherche (ANR) under the project ANR-23-P012-0013-06 (MoreMedDiet), and under the France 2030 programme, reference ANR-22-PESN-0006, project TRACIA. The PRIMA programme is supported by Horizon~2020, the European Union's Framework Programme for Research and Innovation. 

\section*{Data Availability}
The code base for the SSI wallet and the blockchain agent are provided in open-source on Github.
\begin{itemize}
    \item MMD-SSI Wallet: \url{https://github.com/SSI-HF-Research/mmd-wallet}
    \item MMD Blockchain Agent: \url{https://github.com/SSI-HF-Research/mmd-blockchain-agent}
\end{itemize}

\end{document}